\documentstyle[12pt]{article}
\begin{document}
\begin {flushright}
{\small SFU-HEP-04-96} \\
{Revised 07-96}
\end{flushright}
\vspace{0.2cm}

\begin{center}
{\bf String Theory In Curved Space-Time} \\
\vspace{0.2cm}
\end{center}
\vspace{0.5cm}
\begin{center}
K.S.Viswanathan\footnote{Electronic address: kviswana@sfu.ca} \\ 
Department of Physics \\
Simon Fraser University \\
Burnaby,B.C., Canada V5A 1S6 \\
\vspace{0.2cm}
and \\
\vspace{0.2cm}
R.Parthasarathy\footnote{Electronic address: sarathy@imsc.ernet.in}\\
The Institute of Mathematical Sciences \\
C.P.T.Campus, Taramani Post \\
Madras 600 113, India \\
\vspace{0.2cm}
\end{center}

\begin{abstract}
Intrinsic and extrinsic geometric properties of string world sheets in 
curved space-time background are explored. In our formulation, the only 
dynamical degrees of freedom of the string are its immersion 
coordinates. Classical equation of motion and the space-time 
energy-momentum tensor of the string  are obtained. The equations of 
motion for the extrinsic curvature action are second order for  the scalar 
mean curvature of the world sheet. 1-loop divergent terms are 
calculated using the background field method. Asymptotic freedom of the 
extrinsic curvature coupling is established.
\end{abstract}

PACS number(s): 11.25.-w, 11.25.Pm

\newpage

\subsection{INTRODUCTION}

\vspace{0.5cm}

String theory in curved space-times is an exciting subject and in fact has 
been investigated by many authors [ see \cite{vega} for a recent review] as a 
framework to study the physics of gravitation in the context of string 
theory. In most investigations so far, the starting point for open or 
closed bosonic strings propagating in a D-dimensional space-time with 
metric  $h_{\mu\nu}(X)\ ;\ (0\ \leq \ \mu,\nu \ \leq \ D-1)$ is the action
\begin{eqnarray}
S &=& \frac{1}{2\pi{\alpha}'}\int d\sigma d\tau \sqrt{g}\,
g_{\alpha\beta}(\sigma,\tau)h_{\mu\nu}(X)
{\partial}_{\alpha}X^{\mu}(\sigma,\tau){\partial}_{\beta}X^{\nu}
(\sigma,\tau).
\end{eqnarray} 
Here $g_{\alpha\beta}(\sigma,\tau)\ (\alpha,\beta\ =\ 1,2)$ is the
metric on the world sheet. ${\alpha}'$ is the string tension. The metric 
$h{_{\mu\nu}}$ may be taken as either a fixed background or as 
dynamically generated by the string. The dynamical variables of the 
theory described by (1) are the immersion coordinates 
$X^{\mu}(\sigma,\tau)$ and the world sheet metric $g_{\alpha\beta}$. Let 
us recall that the above action is obtained from the Nambu-Goto (N-G) action
\begin{eqnarray}
S &=& \frac{1}{2\pi{\alpha}'}\int \sqrt{g}\, d\sigma d\tau ,
\end{eqnarray}
which in turn can be written as  
\begin{eqnarray}
S &=& \frac{1}{2\pi{\alpha}'}\int (\sqrt{g} + 
\lambda^{\alpha\beta}(\partial_{\alpha}X^{\mu}{\partial}_{\beta}
X^{\nu}h_{\mu\nu}  
- g_{\alpha\beta})) d\sigma d\tau,
\end{eqnarray}
and presuming condensation of the Lagrange multiplier field 
$\lambda_{\alpha\beta}$ \cite{poly:book}
\begin{eqnarray}
<\lambda_{\alpha\beta}> &=& const.\sqrt{g}\, g_{\alpha\beta}.
\end{eqnarray}
The reasons for starting with the alternative (1) in string theory are 
well known. The main points being: the integration over $X^{\mu}$ is 
easily done and the full power of the theory of Riemann surfaces can be 
exploited when integrating over the space of all 2-metrics.

In investigating strings in a curved background, it will be shown in this 
paper that it is more convenient to work with the N-G action (2), where 
the only dynamical variables are $X^{\mu}(\sigma,\tau)$ while the metric 
of the world sheet is induced from its immersion in background geometry. 
In this scheme we do not have to contend with the Weyl symmetry of the 2-d 
metric. It will also become evident below that, since the geometric 
properties of the world sheet are described in terms of both its first 
and  second fundamental forms, it is natural to take into account, 
besides the usual N-G action, an action that involves the extrinsic 
geometry of the world sheet. 2-d surfaces that extremize the N-G action 
have $H^{i} = 0$, where $H^{i} (i=1,...,D-2)$ are the scalar mean 
curvatures of the surface. A much wider class of surfaces extremize the 
extrinsic curvature action, thereby, enabling one to investigate the role 
of a very general class of world sheets on the dynamics of strings. 
There exists a 
large body of work \cite{p}-\cite{c} on strings with
extrinsic curvature 
action in Lorentzian space-time with the hope of describing QCD 
strings. The necessity of including such a term for QCD was emphasized in 
\cite{p}. Extrinsic geometry dependent term in the action provides 
rigidity to the strings, while the N-G term only provides tension. 
Rigidity effects are important also in determining the shapes of 
biological membranes \cite{H},\cite{Pelti}.
It is thus natural to investigate the role of extrinsic geometry on the 
properties of strings in curved space-time. String theory has a 
dimensional parameter, i.e., the string tension. Rigid strings have a 
dimensionless coupling which in flat background is asymptotically free. We 
shall see that this coupling is asymptotically free in curved background 
as well.

Although we discuss primarily string world sheet in curved background,
many of the results obtained here may be used in the context of the
(3+1) canonical formulation of gravity. It may be recalled that in the
(3+1) formulation of Arnowitt, Deser and Misner (ADM) \cite{ADM},\cite{MTW}, the
Cauchy problem of general relativity can be tackled by constructing a
space-time $\tilde{M}$ out of a foliation of space-like
hypersurfaces with time coordinate $t$ as a parameter. In this
formalism the quantities needed for the description of the
4-dimensional line element are the lapse function, which measures
distance between nearby hypersurfaces, the shift vector that gives the
relation between the spatial coordinate systems on different
hypersurfaces and the 3-d space metric to enable one to measure
distances in space-like, t = constant slices. In 4-d space-time,
Einstein's theory leads to three momentum
constraints, the Hamiltonian constraint, and six equations of motion.
The hyperbolic form of
Einstein's theory can be obtained \cite{AACY},\cite{AY} by imposition of
a slicing
condition on the space-time. One such condition involves taking t =
constant surfaces to be constant mean curvature hypersurfaces. Thus,
for example, using harmonic Gauss map , one can solve the
constraints in (2+1) gravity \cite{HOKN}-\cite{KSV}. The considerations 
in this paper
offer other possibilities for choosing hypersurfaces that extremize
the extrinsic curvature action. 2-d surfaces of prescribed mean
curvature are studied as solutions of the Einstein constraint
equations on closed manifolds in \cite{Krivan},\cite{Isenberg}.

Earlier studies in string theory in curved background, for example, in 
\cite{Let} - \cite{Lar}, concentrated on 
 the classical equation of motion in order to 
exhibit its nonlinearity.  
While the authors in 
\cite{Let}, \cite{Guv} and \cite{Lar} considered Nambu-Goto action in
curved background, \cite{Bois} and \cite{Car} (in the second
paper) included extrinsic curvature dependent term in the action. In
these studies the first variation of the action giving the classical
equations of motion is examined for various choices of the background
spacetime. 
Some of the classical aspects of strings in curved background
discussed in this paper have also been considered  recently by Capovilla and
Guven \cite{Cap}.  
They have considered the geometry of deformations
of relativistic membranes. 
In particular they have derived the second
variation of the NG action in a general curved space and that of the
extrinsic curvature action in a background Minkowski spacetime. In
this paper we are interested in both the classical and the quantum
aspects of strings in curved space. In this paper we calculate the second
variation of
both the NG and the extrinsic curvature actions in arbitrary curved
space-time which is then used to calculate the renormalization of both the
dimensionless extrinsic curvature coupling and the background metric.
Furthermore in the path integral formalism adopted here where the
dynamical variables of the theory are the string's immersion
coordinates only, it is shown here explicitly that the volume of the
two dimensional diffeomorphism group is cancelled by the volume of the
space of tangential fluctuations of the world sheet of the string.
This demonstration for onshell amplitudes turns out to be nontrivial.

This paper is organized as follows. In section {\bf 0.2}, we consider the 
classical properties of both the N-G and the extrinsic curvature actions 
in a curved space-time. We derive their equations of motion and discuss 
possible solutions. The space-time 
energy momentum tensor $T^{\mu\nu}$ of the string is derived in
section {\bf 0.3} and in the appendix we establish that it is covariantly
conserved. In 
section {\bf 0.4} we discuss the 1-loop divergences of the theory by 
integrating over $X^{\mu}$. As a consequence of reparametrization 
invariance only the fluctuations normal to the world sheet are 
dynamical. The longitudinal fluctuations are zero modes and their volume 
cancels the volume of the diffeomorphism group of the world sheet. 
It will be seen that a divergent Euler characteristic term 
appears at the 1 loop level while such a term was previously known to appear
only at the 2 loop level in 
the conventional approach \cite{poly:book},\cite{Fradkin}, where the 
dynamical degrees of freedom are the immersion coordinates and the metric 
of the world sheet. The condition 
for 1 loop finiteness is also derived.

\vspace{1.0cm}

\subsection{N-G AND EXTRINSIC CURVATURE  
            ACTIONS IN CURVED SPACE-TIME}
\vspace{0.5cm}            

In the previous section we discussed the N-G action. If the metric 
$h_{\mu\nu}$ of space-time is taken as dynamical, then, we must include 
the Hilbert-Einstein action 
\begin{eqnarray}
S_{H-E} &=&  - \frac{1}{8\pi G} \int \tilde{R} \sqrt{h}\, d^DX.
\end{eqnarray}
We denote the scalar curvature of space-time $\tilde{M}$ by
$\tilde{R}$. We use the notation whereby geometric quantities of 
space-time will be denoted with a tilde and the world sheet quantities 
will be unadorned.    

In order to understand rigid strings, it is necessary to recall the
structure equations for immersed surfaces in $\tilde{M}$. First, the
equation of Gauss
\begin{eqnarray}
{\partial}_{\alpha}{\partial}_{\beta}X^{\mu} +
{\tilde{\Gamma}}^{\mu}_{\nu\rho}
{\partial}_{\alpha}X^{\nu}{\partial}_{\beta}X^{\rho} - 
{\Gamma}^{\gamma}_{\alpha\beta}{\partial}_{\gamma}X^{\mu} &=& 
H^i_{\alpha\beta} N^{i\mu},
\end{eqnarray} 
defines the second fundamental form $H^i_{\alpha\beta}$ (i =
1,2,....D-2). In (6) ${\tilde{\Gamma}}^{\mu}_{\nu\rho}$ is the
connection in $\tilde{M}$ determined by the metric $h_{\mu\nu}$, while
${\Gamma}^{\gamma}_{\alpha\beta}$ is the connection coefficient on
string world sheet $M$ determined by the induced metric
$g_{\alpha\beta}$ on $M$. $N^{i\mu}$ (i = 1,2,...D-2) are the (D-2)
normals to the string world sheet at $X^{\mu}(\sigma,\tau)$. We choose
the normalization
\begin{eqnarray}
N^{i\mu} N^{j\nu} h_{\mu\nu} &=& {\delta}_{ij} \ \epsilon(i),
\end{eqnarray}
and 
\begin{eqnarray}
{\partial}_{\alpha}X^{\mu} N^{i\nu} h_{\mu\nu} &=& 0, \ \ \ \ \ (i\
=\ 1,2,...D-2).
\end{eqnarray}
$\epsilon(i)\ =\ \pm 1$ depending on whether the world sheet $M$ has
an induced metric which is Riemannian or pseudo-Riemannian. For string
theory, the world sheet metric has Lorentzian signature and hence
$\epsilon(i)$ in (7) is +1 for i = 1,2,...D-2. In the context of
canonical (3+1) gravity, the hypersurface has induced metric which is
Riemannian and so $\epsilon(i)$ = -1. Note that (6) may be expressed
in compact form as
\begin{eqnarray}
{\nabla}_{\alpha} {\nabla}_{\beta} X^{\mu} &=&
H^i_{\alpha\beta} N^{i\mu}, \nonumber
\end{eqnarray}
where ${\nabla}_{\alpha}$ is the symbol for covariant
derivative. Note that ${\nabla}_{\beta}X^{\mu}\ \equiv \
{\partial}_{\beta} X^{\mu}$ is a world-vector as well as a 2-vector and
hence needs both connections ${\tilde{\Gamma}}^{\mu}_{\nu\rho}$ and
${\Gamma}^{\gamma}_{\alpha\beta}$ to define its covariant derivative.

Next, we have the equation of Weingarten
\begin{eqnarray}
{\partial}_{\alpha}N^{i\mu} +
{\tilde{\Gamma}}^{\mu}_{\nu\rho}{\partial}_{\alpha}X^{\nu} N^{i\rho}- 
A^{ij}_{\alpha} N^{j\mu} &=& -\epsilon(i)
H^{i\beta}_{\alpha}{\partial}_{\beta}X^{\mu}.
\end{eqnarray}
In (9)
\begin{eqnarray}
A^{ij}_{\alpha} &=& \epsilon(i) \left(
N^{j\nu}{\tilde{\nabla}}_{\alpha} N^{i\mu}\right)
h_{\mu\nu}.
\end{eqnarray}
\begin{eqnarray}
{\tilde{\nabla}}_{\alpha} N^{i\mu} &=&
{\partial}_{\alpha}N^{i\mu} +
{\tilde{\Gamma}}^{\mu}_{\nu\rho}{\partial}_{\alpha}X^{\nu} N^{i\rho}.
\end{eqnarray}
$A^{ij}_{\alpha}$ is the 2-d gauge field or gauge connection in the
normal bundle.

We shall use the notation whereby we covariantize on all indices. $\mu $
is the world index, $\alpha \ , \beta $, the 2-dimensional world
sheet index, and i,j the `internal' indices of the normal bundle. Thus
\begin{eqnarray}
{\nabla}_{\alpha} N^{i\mu}&\equiv & {\partial}_{\alpha}
N^{i\mu} + {\tilde{\Gamma}}^{\mu}_{\nu\rho} {\partial}_{\alpha}
X^{\nu} N^{i\rho} - A^{ij}_{\alpha} N^{j\mu}.
\end{eqnarray}
Note that in D-dimensional space-time $A^{ij}_{\alpha}\ =\
-A^{ji}_{\alpha}$ are 2-dimensional gauge fields with the gauge group 
$SO(D-2)$.

In addition to the equations of Gauss and Weingarten, there are three
key equations we need in the sequel. These are the equations of
Gauss, Codazzi, and Ricci. These are well known and we merely give
these below for completeness \cite{Eisenhart},\cite{Y-K}.

{\bf Equation of Gauss:}

\begin{eqnarray}
{\tilde{R}}_{\mu\nu\rho\sigma} {\partial}_{\alpha}X^{\mu}
{\partial}_{\beta} X^{\nu} {\partial}_{\gamma}X^{\rho}
{\partial}_{\delta}X^{\sigma} &=& R_{\alpha\beta\gamma\delta}
\nonumber \\ 
& & +\sum_{i}\epsilon(i)\left( H^i_{\beta\gamma} H^i_{\alpha\delta} -
H^i_{\beta\delta}H^i_{\alpha\gamma}\right),
\end{eqnarray}
where $R_{\alpha\beta\gamma\delta}$ is the curvature of the string
world sheet.

{\bf Equation of Codazzi:}

\begin{eqnarray}
{\nabla}_{\alpha} H^i_{\beta\gamma} -
{\nabla}_{\beta}H^i_{\alpha\gamma} &=& \epsilon(i)
{\tilde{R}}_{\mu\nu\rho\sigma}{\partial}_{\alpha}X^{\mu}{\partial}_
{\beta}X^{\nu}{\partial}_{\gamma}X^{\sigma} N^{i\rho},
\end{eqnarray}
and

{\bf Equation of Ricci:}

\begin{eqnarray}
{\tilde{R}}_{\mu\nu\rho\sigma}{\partial}_{\alpha}X^{\mu} 
{\partial}_{\beta}X^{\nu}N^{j\rho} N^{i\sigma}&=& F^{ij}_{\alpha\beta}
  + H^{i\gamma}_{\alpha} H^j_{\gamma\beta} - H^{i\gamma}_{\beta}
H^j_{\gamma\alpha},
\end{eqnarray}
where
\begin{eqnarray}
F^{ij}_{\alpha\beta} &=& {\partial}_{\alpha}A^{ij}_{\beta} -
{\partial}_{\beta} A^{ij}_{\alpha} + A^{ik}_{\alpha} A^{kj}_{\beta} -
A^{jk}_{\alpha} A^{ki}_{\beta}.
\end{eqnarray}
In (12) we have used the notation
\begin{eqnarray}
{\nabla}_{\alpha}H^i_{\beta\gamma} &=&
{\partial}_{\alpha}H^i_{\beta\gamma} - {\Gamma}^{\delta}_{\alpha\beta}
H^i_{\delta\gamma} 
  - {\Gamma}^{\delta}_{\alpha\gamma}H^i_{\beta\delta} -
A^{ij}_{\alpha} H^j_{\beta\gamma}.
\end{eqnarray}

In general these are the only relations between the curvature tensor
of the space-time and the extrinsic curvature of the world sheet. In
the case when $M$ is a hypersurface in space-time there is only one
normal and so $A^{ij}_{\alpha}\ \equiv \ 0$. Further relations may be
obtained when space-time metric is written in terms of, say, the
isothermal coordinates \cite{MTW},\cite{LPPT}. The choice of the
extrinsic curvature
action for strings is motivated from the relation that follows from
(13) when ${\tilde{R}}_{\mu\nu\rho\sigma} \ =\ 0$ ($\tilde{M}$ is
Minkowski). In this case,

\begin{eqnarray}
R &=& \sum_{i}\epsilon(i) \left( H^i_{\alpha\beta}H^{i\alpha\beta} -
(H^{i\alpha}_{\alpha})^2 \right),
\end{eqnarray}
where $R$ is the scalar curvature of the world sheet. In 2-dimensions

\begin{eqnarray}
\frac{1}{4\pi} \int d\sigma d\tau \sqrt{g}\, R &=& \chi,
\end{eqnarray}
where $\chi$ is the Euler characteristic of the two-dimensional manifold.
On the
other hand each term on the right in (18) is separately
reparameterization invariant and hence can be used as the extrinsic
curvature action. For convenience, we take this as (subscript R for rigid
strings)

\begin{eqnarray}
S_R &=& \frac{1}{{\alpha}_0} \int \sqrt{g}\, H^2 d\sigma d\tau,
\end{eqnarray}
where

\begin{eqnarray}
H^2 &=& \sum_{i} {\left( \frac{1}{2}H^{i\alpha}_{\alpha}\right)}^2.
\end{eqnarray}

We continue to take (20) as the extrinsic curvature action for rigid
strings in curved background [From now on we shall drop the factor $\epsilon(i)$
for convenience. In fact for strings all normals will be space-like
and $\epsilon(i)\ =\ +1$]. ${\alpha}_0$ in (20) is a dimensionless
coupling. In a flat background, it is known that this coupling is
asymptotically free \cite{p},\cite{v}. Note that if we considered an action

\begin{eqnarray}
S&=& \int
{\tilde{R}}_{\mu\nu\rho\sigma}{\partial}_{\alpha}X^{\mu}{\partial}_
{\beta}X^{\nu} {\partial}_{\gamma}X^{\rho}{\partial}_{\delta}X^
{\sigma} g^{\beta\gamma} g^{\alpha\delta} \sqrt{g}\, d\sigma d\tau,
\end{eqnarray}
it will not take into account extrinsic curvature or bending rigidity
effects although such terms will probably arise in the effective
action starting from the N-G term.   

In Minkowski space-time (20), besides being reparameterization
invariant on the world sheet, is scale invariant under $X^{\mu} \
\rightarrow \ \lambda X^{\mu}$. Although not usually emphasized in the
literature, the action (20) is actually invariant under space-time
conformal transformations \cite{Blaschke}-\cite{Willmore}. 
It is not too difficult to show
that (20) is invariant under 4-translations, scale transformation,
Lorentz transformations and special conformal transformations \cite{Pokorski}
\begin{eqnarray}
X'^{\mu} &=& \frac{X^{\mu} - b^{\mu} X^2}{1 - 2 b\cdot X + b^2 X^2},
\end{eqnarray}
where $b^{\mu}$ is a constant D-vector. Invariance under the first three  is
manifest. It is readily verified that under infinitesimal special
conformal transformation, (20) is transformed into a total divergence
and hence either vanishes for closed surfaces (without boundary) or
turns into a boundary contribution.

Let us concentrate on the properties of (20) in curved background. In
order to appreciate the nature of the string world sheet that satisfies
the equation of motion of (20), it is instructive to start with the
N-G action (2). A straightforward computation shows that the equation
of motion of (2) can be written as
\begin{eqnarray}
\sum_{i} N^{i\mu} H^i &=& 0.
\end{eqnarray}
This is  equivalent to $H^i\ =\ 0,\ i\ =\ 1,2,...D-2$. In
other words surfaces with zero mean curvature,  i.e., minimal
surfaces, extremize the N-G action. This conclusion is 
independent of whether the background space-time is flat or curved. Note 
that the equation of motion for the N-G action is algebraic in the scalar 
mean curvature. In
flat space-time it reduces to 
\begin{eqnarray}
({\partial}^2_{\tau} - {\partial}^2_{\sigma}) X^{\mu} &=& 0,
\end{eqnarray}
upon using isothermal coordinates on the string world sheet. In curved
background  it reads
\begin{eqnarray}
g^{\alpha\beta} {\nabla}_{\alpha}{\partial}_{\beta} X^{\mu}
&=& 0,
\end{eqnarray}
where ${\nabla}_{\alpha}{\partial}_{\beta}X^{\mu}$ is
defined in  (6).  

In the path integral formulation, classical solutions provide a starting
point for non-perturbative calculations. Thus one can integrate over
minimal surfaces for N-G action. In fact the present authors have
carried out such a calculation for minimal and harmonic ($H^i\ =\
constant$) surfaces for rigid strings in a flat background space-time
and obtained a modified Coulomb gas
picture of QCD effective string action \cite{vp}. This demonstration 
\cite{vp} however utilized the notion of the Gauss map from the world 
sheet into the Grassmannian manifold.

Next, consider the equation of motion of rigid string action in curved
space-time. Surprisingly, the derivation is rather involved in this
case. The special case of an hypersurface in D-dimensions has been discussed
in \cite{W-J}. We sketch below the main steps.  

Starting from (20) we have, under $X^{\mu}\ \rightarrow \ X^{\mu}\ +\
\delta X^{\mu}$,

\begin{eqnarray}
\delta S_R &=& \frac{1}{{\alpha}_0}\left( \frac{1}{2}\int \sqrt{g}\,
g^{\alpha\beta}\delta g_{\alpha\beta} H^2 d\sigma d\tau + 2\int H^i\delta H^i
\sqrt{g}\, d\sigma d\tau \right).
\end{eqnarray}

A simple calculation yields
\begin{eqnarray}
\delta g_{\alpha\beta} &=& {\nabla}_{\alpha}\left( \delta
X^{\mu} {\partial}_{\beta}X^{\nu} h_{\mu\nu}\right) +
{\nabla}_{\beta}\left(\delta X^{\mu} {\partial}_{\alpha}
X^{\nu} h_{\mu\nu} \right) \nonumber \\  
 & & - 2\delta X^{\mu} h_{\mu\nu} N^{i\nu} H^i_{\alpha\beta}.
\end{eqnarray}
 
\begin{eqnarray}
\delta N^{i\mu} + \delta X^{\sigma} N^{i\nu} \Gamma^{\mu}_{\sigma\nu}
= - {\partial}_{\alpha}X^{\mu} \left[{\nabla}^{\alpha}(N^{i\nu} 
\delta X^{\sigma}h_{\nu\sigma}) 
+H^{i\alpha\beta}{\partial}_{\beta}X^{\nu} \delta X^{\rho} 
h_{\nu\rho}\right].
\end{eqnarray}
  
\begin{eqnarray}
\delta H^i &=& -{\nabla}_{\alpha}\left( H^{i\alpha\beta}
\delta X^{\mu} {\partial}_{\beta}X^{\nu} h_{\mu\nu}\right) + \left(
{\nabla}_{\alpha} H^{i\alpha\beta}\right){\partial}_{\beta}
X^{\nu} \delta X^{\mu} h_{\mu\nu} \nonumber \\
 & & + H^{i\alpha\beta}H^j_{\alpha\beta}N^{j\nu}\delta X^{\mu}h_{\mu\nu}
+ \frac{1}{2} g^{\alpha\beta} \delta H^i_{\alpha\beta}.
\end{eqnarray}
 
\begin{eqnarray}
g^{\alpha\beta}H^i \delta H^i_{\alpha\beta} &=&
{\nabla}^{\alpha}\left( ({\nabla}_{\alpha}\delta
X^{\mu})N^{i\nu}h_{\mu\nu}H^i-\delta
X^{\mu}h_{\mu\nu}{\nabla}_{\alpha}(N^{i\nu}H^i)\right)
\nonumber \\
& & + g^{\alpha\beta}H^i
{\tilde{R}}_{\mu\nu\rho\sigma}{\partial}_{\alpha}X^{\nu}{\partial}_
{\beta}X^{\rho}\delta X^{\sigma}N^{i\mu} \nonumber \\
& & +\delta X^{\mu}h_{\mu\nu}[ H^i\left( -({\nabla}^{\alpha}
H^{i\gamma}_{\alpha}){\partial}_{\gamma}X^{\nu} - H^iH^j_{\alpha\beta}
H^{i\alpha\beta}N^{j\nu}\right) \nonumber \\
& & - 2H^{i\alpha\gamma}({\partial}_{\gamma}X^{\nu}){\nabla}_
{\alpha} H^i + N^{i\nu}{\nabla}^2 H^i ].
\end{eqnarray} 
 
In (31), ${\tilde{R}}_{\mu\nu\rho\sigma}$ is the curvature tensor of
the background space-time defined by

\begin{eqnarray} 
{\tilde{R}}^{\mu}\;_{\nu\rho\sigma}&=&{\partial}_{\rho}{\tilde{\Gamma}}
^{\mu}_{\nu\sigma}-{\partial}_{\sigma}{\tilde{\Gamma}}^{\mu}_{\nu\rho}
+{\tilde{\Gamma}}^{\lambda}_{\nu\sigma}{\tilde{\Gamma}}^{\mu}_{\rho
\lambda} 
- {\tilde{\Gamma}}^{\lambda}_{\nu\rho}{\tilde{\Gamma}}^{\mu}_{\sigma
\lambda} .
\end{eqnarray}

Let us note from equation(29) that $\delta N^{i\mu}$ is not a covariant 
vector, but that $\delta N^{i\mu} + \delta X^{\sigma} 
\Gamma^{\mu}_{\sigma\nu} N^{i\nu}$ is. 
Using (28)-(31) in (27) 
 one finds
 
\begin{eqnarray}
\delta S_R&=&\frac{1}{{\alpha}_0}\int \{\nabla^{\alpha} [(H^2 
g_{\alpha\beta} - 2H^i H^i_{\alpha\beta})\delta X^{\mu}{\partial}^{\beta} 
X^{\nu} h_{\mu\nu}  \nonumber \\
& & + ({\nabla}_{\alpha} \delta X^{\mu}) 
N^{i\nu}h_{\mu\nu}H^i - {\delta}X^{\mu} h_{\mu\nu} 
{\nabla}^{\alpha}(N^{i\nu} H^{i}) ] \nonumber \\
& & +  [\left( 
-2({\nabla}^{\alpha}H^i)H^i +
H^i{\nabla}_{\beta}H^{i\alpha\beta}\right)
{\partial}_{\alpha}X^{\nu} \delta X^{\mu} h_{\mu\nu}  \nonumber \\
& & + \left({\nabla}^2 H^i - 2 H^2 H^i + H^j
H^j_{\alpha\beta}H^{i\alpha\beta}\right) \delta X^{\mu} N^{i\nu}
h_{\mu\nu}  \nonumber \\
& & - g^{\alpha\beta} H^i
{\partial}_{\alpha}X^{\nu}{\partial}_{\beta}X^{\rho}N^{i\mu}\delta
X^{\sigma} {\tilde{R}}_{\mu\nu\rho\sigma}]\} \sqrt{g}\, d\sigma d\tau.
\end{eqnarray}
Equations of motion can be obtained by decomposing $\delta X^{\mu}$ into
normal and tangential variations of the world sheet. Accordingly, we
write

\begin{eqnarray}
\delta X^{\mu} &=& {\xi}_j N^{j\mu} +
{\partial}_{\alpha}X^{\mu} {\xi}^{\alpha}.
\end{eqnarray}
Then it follows from (33) that the equation of motion for normal
variations is

\begin{eqnarray}
{\nabla}^2 H^i - 2 H^2 H^i + H^j H^{j\alpha\beta}
H^i_{\alpha\beta} 
- g^{\alpha\beta} H^j
{\tilde{R}}_{\mu\nu\rho\sigma}{\partial}_{\alpha}X^{\nu}{\partial}_
{\beta}X^{\rho} N^{j\mu} N^{i\sigma} = 0, 
\end{eqnarray}
where i = 1,2,...D-2.
The tangential variations yield

\begin{eqnarray}
\sum_{i} H^i [ {\nabla}_{\beta} H^{i\alpha\beta} -
2{\nabla}^{\alpha}H^i 
-{\tilde{R}}_{\mu\nu\rho\sigma}{\partial}_{\gamma}X^{\nu}{\partial}
_{\delta}X^{\rho} g^{\gamma\delta}{\partial}^{\alpha}X^{\sigma}
N^{i\mu} ] & =&  0.
\end{eqnarray}
The equations of motion for the action that is a sum of the N-G and
extrinsic curvature actions is easily seen to be

\begin{eqnarray}
{\nabla}^2H^i-2H^i(H^2 + k) +
H^jH^{j\alpha\beta}H^i_{\alpha\beta} \nonumber \\ 
- g^{\alpha\beta}H^j{\tilde{R}}_{\mu\nu\rho\sigma}
{\partial}_{\alpha}X^{\nu}{\partial}_{\beta}
X^{\rho}N^{j\mu}N^{i\sigma} &=& 0,
\end{eqnarray}
where $k\ =\ -\frac{{\alpha}_0}{2\pi{\alpha}'}$. Equation (36) is
however unchanged.  
In (35)

\begin{eqnarray}
{\nabla}^2 &\equiv & g^{\alpha\beta}
{\nabla}_{\alpha}{\nabla}_{\beta}; \nonumber \\
{\nabla}_{\alpha}H^i &=& {\partial}_{\alpha}H^i -
A^{ij}_{\alpha}H^j.
\end{eqnarray}
Equation (36) is actually not an equation of motion. It is just the 
structure equation of Codazzi (14) once contracted.
This is just a consequence of reparametrization 
invariance of the action. Thus the surfaces with mean curvature $H^i$
satisfying (35) extremize the rigid string action in curved
background. It is worth remarking here that the extrinsic curvature action 
contains fourth derivative operator acting on $X^{\mu}$. Neverthless its 
equation of motion is expressible as a second order non linear 
partial differential equation in the mean scalar curvature $H^{i}$ of the 
world sheet. Once one has a solution of equation (35) for $H^{i}$ then 
one can solve for immersion coordinates $X^{\mu}$ from equation (6).
Solution of equation of motion for N-G strings in
cosmological space-times \cite{vega} reveals stable, dual to stable and
unstable solutions. In the language employed here, minimal surfaces
are still solutions to N-G string in curved space-times. Hence the
result of \cite{vega} indicates that not all minimal surfaces are stable
immersions in space-time. It would thus be interesting to investigate
similar stability considerations for rigid strings. In flat space-time,
solutions to (35) in 3-dimensional Euclidean space are referred to as
Wilmore surfaces \cite{Willmore} and have been investigated in the context of
biological membranes in \cite{Kholodenko}. This equation takes the form (after
using (13))
\begin{eqnarray}
{\nabla}^2H + 2H(H^2 - \frac{R}{2}) &=& 0,
\end{eqnarray}
where $R$ is the scalar curvature of the world sheet. A sphere
satisfies this equation with constant $H$. Solutions with non-zero genus
are known to exist for (39) \cite{Willmore}. Note that (39) is a nonlinear wave
equation for $H^i$ in the context of string theory.

The special case of a hypersurface in curved space-time is worth
considering. In this case the indices $i,j$ take only one value and
$\alpha , \beta$ take values 1,2,3. Then (35) reads
 
\begin{eqnarray}
{\nabla}^2H-2H^3+HH_{\alpha\beta}H^{\alpha\beta}-H
{\tilde{R}}_{\mu\nu\rho\sigma}{\partial}_{\alpha}X^{\nu}
{\partial}_{\beta}X^{\rho}g^{\alpha\beta}N^{\mu}N^{\sigma}& =& 0.
\end{eqnarray}
Using (13) to eliminate $H_{\alpha\beta}H^{\alpha\beta}$ from (40) we
find

\begin{eqnarray}
{\nabla}^2H+2H^3+H [
{\tilde{R}}_{\mu\nu\rho\sigma}{\partial}_{\alpha}X^{\mu}{\partial}_
{\beta}X^{\nu}{\partial}_{\gamma}X^{\rho}{\partial}_{\delta}X^{\sigma}
g^{\alpha\gamma}g^{\beta\delta}  \nonumber \\
-R - {\tilde{R}}_{\mu\nu\rho\sigma}{\partial}_{\alpha}X^{\nu}
{\partial}_{\beta}X^{\rho} g^{\alpha\beta} N^{\mu} N^{\sigma}] &=& 0.
\end{eqnarray}
 
In \cite{W-J} the term
${\tilde{R}}_{\mu\nu\rho\sigma}{\partial}_{\alpha}X^{\mu}{\partial}_
{\beta}X^{\nu}{\partial}_{\gamma}X^{\rho}{\partial}_{\delta}X^
{\sigma} g^{\alpha\gamma}g^{\beta\delta}$ is replaced by $\tilde{R}$,
which in their notation presumably  is the scalar curvature of $\tilde{M}$,
which is
given by ${\tilde{R}}_{\mu\nu\rho\sigma}h^{\mu\rho}h^{\nu\sigma}$. This
would imply that ${\partial}_{\alpha}X^{\mu}{\partial}_{\gamma}X^{\rho}
g^{\alpha\gamma} = h^{\mu\rho}$ which is, of course, wrong [see 
equation (57) below].

In string theory, it would be most interesting to investigate
solutions to (35) in cosmological backgrounds. Let us however restrict
here to de Sitter or anti-de Sitter space-times, i.e., space-times of
constant curvature. We can write 
\begin{eqnarray}
{\tilde{R}}_{\mu\nu\rho\sigma} &=& K( h_{\mu\rho}h_{\nu\sigma} -
h_{\mu\sigma}h_{\nu\rho}).
\end{eqnarray}
The equation of motion (35) reads (we choose $\epsilon(i)\ =\ 1$)
\begin{eqnarray}
{\nabla}^2H^i - 2H^i (H^2 + K) + H^j
H^j_{\alpha\beta}H^{i\alpha\beta} &=& 0.
\end{eqnarray}
It would be of interest to analyze solutions to (43).

Locally conserved currents arising as a consequence of the invariance of 
the extrinsic curvature action under local space-time conformal 
transformations may be readily obtained from the total divergence terms 
in (33). Let us discuss, in particular only the string's space-time 
momenta. It is easily seen to be given by 

\begin{eqnarray}
(P^{\alpha}_{\mu})_{R} &=& \frac{\sqrt{g}}{\alpha_{0}}\,[(H^2 
g^{\alpha\beta} - H^i H^{i\alpha\beta}) 
{\partial}_{\beta}X^{\nu}h_{\mu\nu} - 
(\nabla^{\alpha}H^i)N^{i\nu}h_{\mu\nu}].
\end{eqnarray}
We notice that $(P^{\alpha}_{\mu})_{R}N^{i\mu}=  
-\frac{\sqrt{g}}{\alpha_{0}}(\nabla^{\alpha}H^{i})$ is the component of 
string momenta normal to its world sheet. This measures the bending 
energy of the rigid string. In comparison the space-time momentum of the 
N-G string is given by
\begin{eqnarray}
(P^{\alpha}_{\mu})_{N-G} &=& \frac{\sqrt{g}}{2\pi \alpha'} 
g^{\alpha\beta}({\partial}_{\beta}X^{\nu})h_{\mu\nu}
\end{eqnarray}
which has no component normal to the world sheet.

\subsection{STRING ENERGY-MOMENTUM TENSOR}

The string's space-time energy-momentum tensor $T^{\mu\nu}$ is obtained
by taking the functional derivative of the action with respect 
to the metric
$h_{\mu\nu}$ at the space-time point X. The calculation nevertheless
turns out to be rather long. We quote here the result.
    
\begin{eqnarray}
\delta S_R &=& \frac{1}{2}\int \sqrt{h}\, T^{\mu\nu} \delta h_{\mu\nu}\,
d^DX,
\end{eqnarray}
with 
\begin{eqnarray}
T^{\mu\nu}(X)&=& \frac{1}{{\alpha}_0}\int d\sigma d\tau\,
\frac{{\delta}^D(X-X(\sigma,\tau))}{\sqrt{h}}\{
\sqrt{g}\,{\partial}_{\alpha}X^{\mu}{\partial}_{\beta}
X^{\nu} \nonumber \\
& & (H^2g^{\alpha\beta} - H^iH^{i\alpha\beta}) \nonumber \\
 & & - \frac{1}{2}\sqrt{g}\,g^{\alpha\beta}{\nabla}_{\beta}
H^i (N^{i\nu}{\partial}_{\alpha}X^{\mu}+N^{i\mu}{\partial}_{\alpha}
X^{\nu}) \} \nonumber \\
& & -\frac{{\nabla}_{\rho}}{2{\alpha}_0}\int d\sigma d\tau\,
\frac{{\delta}^D(X - X(\sigma,\tau))}{\sqrt{h}}\sqrt{g}\,g^{\alpha\beta}
H^i [{\partial}_{\beta}X^{\mu}(N^{i\nu}{\partial}_{\alpha}X^{\rho}
\nonumber \\
& & -
N^{i\rho}{\partial}_{\alpha}X^{\nu})+{\partial}_{\beta}X^{\nu}
(N^{i\mu}{\partial}_{\alpha}X^{\rho}-N^{i\rho}{\partial}_{\alpha}
X^{\mu})].  
\end{eqnarray}   
Note that $X$ in (47) is just a
space-time point, whereas $X(\sigma,\tau)$ stands for the string
dynamical variables. One sees from the D-dimensional Dirac delta
function in (47) that $T^{\mu\nu}(X)$ vanishes unless $X$ is exactly
on the string world-sheet. It is instructive to compare (47) with the
energy-momentum tensor of the N-G strings.
\begin{eqnarray}
T^{\mu\nu}_{N-G}(X) &=& \frac{1}{{\alpha}'}\int d\sigma d\tau\,
\frac{{\delta}^D(X-X(\sigma,\tau))}{\sqrt{h}} \sqrt{g}\,
g^{\alpha\beta}{\partial}_{\alpha}X^{\mu}{\partial}_{\beta}X^{\nu}.
\end{eqnarray}
Note that the flux of $T^{\mu\nu}_{N-G}$ along the normal direction
$T^{\mu\nu}_{N-G}\ N^i_{\nu}(X) \ =\ 0$. This is no longer true for
rigid strings as can be seen from (47). This is consistent with the 
remarks in the introduction that bending of the strings costs energy.

It can be verified, although the calculations are rather long, that
\begin{eqnarray}
{\nabla}_{\mu} T^{\mu\nu} &=& 0
\end{eqnarray} 
when the equation of motion are taken into account. Demonstration of
(49) is sketched in the Appendix.   

\vspace{1.0cm}

\subsection{1-LOOP QUANTUM EFFECTS}

We now discuss 1-loop quantum effects for strings in background 
space-time. In the conventional treatment, the N-G action is taken in the 
form (1) in which the dynamical variables are $g_{\alpha\beta}$ and 
$X^{\mu}$ and in this form the action is regarded as a non linear 
$\sigma$ model \cite{poly:book}. We show in this section that 
it is more convenient to perform computations by regarding the world 
sheet as an immersed surface and the metric on it as induced from the 
background space-time. The dynamical fields are just the immersion 
coordinates $X^{\mu}$. It will become clear that in this scheme not only 
the metric of the world sheet but also its extrinsic geometry play a key 
role in the dynamics.
The partition function for the string is given by
\begin{eqnarray}
{\bf Z} &=& \int \frac{{\cal D}X^{\mu}}{Vol}\, e^{-\frac{1}{2\pi{\alpha}'}
\int \sqrt{g}\, d\sigma d\tau} , 
\end{eqnarray}
where $Vol$ is the volume of the 2-d diffeomorphism group.
To calculate  1-loop effects  we need the second variation of the action. 
We now evaluate this on-shell, i.e. for minimal surfaces. As in (34) we 
decompose the fluctuations $\delta X^{\mu}$ into normal and tangential 
components. For the N-G action the first variation of the action can be 
written as
 
\begin{eqnarray}
\delta S &=& - \frac{1}{\pi{\alpha}'} \int \sqrt{g}\, H^i \xi_{i} d\sigma 
d\tau .
\end{eqnarray}

To evaluate the second variation on-shell ($H^i=0$), we need only evaluate 
the variation of $H^i$. It is convenient to rewrite (30) and (31) in the 
following form

\begin{eqnarray}
{\delta H}^{i}_{\alpha\beta} &=& ( {\nabla}_{\alpha} {\nabla}_{\beta} 
{\delta}^{i}_{j} - H^{i}_{\alpha\gamma} H^{i\gamma}_{\beta}   \nonumber \\
& & - {\tilde{R}}_{\mu\nu\rho\sigma}N^{i\mu}N^{j\sigma} 
{\partial}_{\alpha}X^{\rho} {\partial}_{\beta}X^{\nu} ) {\xi}_{j} 
\nonumber \\
& & + \left( H^{i\gamma}_{\alpha} {\nabla}_{\beta} + H^{i\gamma}_{\beta} 
{\nabla}_{\alpha} \right) \xi_{\gamma} + ({\nabla}^{\gamma} 
H^{i}_{\alpha\beta}) \xi_{\gamma} .
\end{eqnarray}

In arriving at the result above from (30) and (31), we have made use of 
the equation of Gauss (13). From above we find that $ \delta H^{i}$ can 
be expressed as follows,
\begin{eqnarray}
\delta H^{i} &=& \frac{1}{2} {\bf O}^{i}_{j} {\xi}^{j} + 
({\nabla}^{\alpha} H^{i}){\xi}_{\alpha},
\end{eqnarray}
where

\begin{eqnarray}
{\bf O}^{i}_{j} &=& {\nabla}^{2} {\delta}^{i}_{j} + H^{i}_{\alpha\beta} 
H^{j\alpha\beta} - {\tilde{R}}_{\mu\nu\rho\sigma} g^{\alpha\beta} 
{\partial}_{\alpha}X^{\nu}{\partial}_{\beta}X^{\rho}N^{i\mu}N^{j\sigma}.
\end{eqnarray}
Substituting from above we can write the second variation of (2) as follows
\begin{eqnarray}
{\delta}^{2} S =  - \frac{1}{2\pi \alpha'} \int \sqrt{g}\, {\xi}^{i} {\bf
O}^{i}_{j} 
{\xi}^{j} d\sigma d\tau.
\end{eqnarray}
Note that only the normal-normal fluctuations appear in the second 
variation when evaluated on-shell. As a consequence  $\int \frac{{\cal 
D}X^{\mu}}{Vol} \rightarrow \int \frac{{\cal D}{\xi}^{\alpha}}{Vol} {\cal 
D}{\xi}^{i} \rightarrow \int {\cal D}{\xi}^{i}$. 

The logarithmic correction to the effective action can now be read off 
from (54) and (55) \cite{Luis}

\begin{eqnarray}
\Gamma^{(1)} &=& \frac{I}{2}\int d\sigma d\tau \sqrt{g}\, \{ 
{\tilde R}_{\lambda\nu\rho\sigma}g^{\alpha\beta}{\partial}_{\alpha}X^{\nu}
{\partial}_{\beta}X^{\rho}N^{i}\,^{\lambda}N^{i}\,^{\sigma}  \nonumber \\
& & - H^{i}_{\alpha\beta}H^{i\alpha\beta}\},
\end{eqnarray}
where 
\begin{eqnarray}
I = \int^{\Lambda} \frac{d^{2}k}{(2\pi)^{2} k^{2}}.
\end{eqnarray}
The covariant derivative $\nabla$ in (54) contains the gauge connection 
$A_{\alpha}^{ij}$. But counter terms dependent on $A_{\alpha}^{ij}$ 
do not arise since they can appear only through $Tr( 
F_{\alpha\beta}F^{\alpha\beta})$ which is a dimension four operator.

Now from Gauss' formula (13) we find

\begin{eqnarray}
H^{i}_{\alpha\beta} H^{i\alpha\beta} &=& 4 H^{2} + R - 
{\tilde{R}}_{\mu\nu\rho\sigma} 
{\partial}_{\alpha}X^{\mu}{\partial}_{\beta}X^{\nu}{\partial}_{\gamma}X^{\rho}
{\partial}_{\delta}X^{\sigma} g^{\beta\gamma} g^{\alpha\delta}.
\end{eqnarray}
$R$ in above is the scalar curvature of the world sheet.
Note that for N-G action we put $H=0$ in the above formula.
Substituting (58) in (56) and making use of the completeness relation 

\begin{eqnarray}
\sum_{i=1}^{D-2} N^{i\lambda} N^{i\sigma} + 
\partial_{\gamma}X^{\lambda} \partial_{\delta}X^{\sigma} 
g^{\gamma\delta} &=& h^{\lambda\sigma},
\end{eqnarray}
we get

\begin{eqnarray}
\Gamma^{(1)} &=&  \frac{I}{2} \int d\sigma d\tau \sqrt{g} \{ 
\tilde{R}_{\nu\rho} {\partial}_{\alpha}X^{\nu}{\partial}_{\beta}X^{\rho} 
g^{\alpha\beta} - R \}
\end{eqnarray}
where $ \tilde{R}_{\nu\rho}$ is the Ricci tensor of space-time. As 
claimed before, the Euler characteristic term appears in 1-loop order. 
This is to be contrasted with the results in \cite{poly:book}, 
\cite{Fradkin}.
Both  divergent terms in (60) can be absorbed in the 
renormalization of the background metric $h_{\mu\nu}$. Thus, defining 
renormalized $h_{\mu\nu}$ by
\begin{eqnarray}
h_{\mu\nu}^R &=& h_{\mu\nu} + \delta h_{\mu\nu},
\end{eqnarray}
we find that
\begin{eqnarray}
\sqrt{g} &=& det({\partial}_{\alpha}X^{\mu} 
{\partial}_{\beta}X^{\nu}(h^{R}_{\mu\nu} - \delta h_{\mu\nu})) \nonumber \\
&\simeq&  \sqrt{g} (1 - \frac{1}{2} g^{\alpha\beta} 
{\partial}_{\alpha}X^{\mu} {\partial}_{\beta}X^{\nu} \delta h_{\mu\nu}).
\end{eqnarray}
Thus, we define 
\begin{eqnarray}
\delta h_{\mu\nu} &=& 2\pi \alpha'I[ \tilde{R}_{\mu\nu} - 
\frac{1}{2} R\, h_{\mu\nu}].
\end{eqnarray}
The divergent Euler characteristic term thus gets absorbed in the 
redefinition of the background metric rather than in Liouville field as 
happens in the conventional procedure \cite{poly:book}.
Thus, for 1-loop finiteness of string theory the necessary condition is
 
\begin{eqnarray}
\tilde{R}_{\mu\nu} - \frac{1}{2} R\, h_{\mu\nu} &=& 0 
\end{eqnarray}
which is different from the Ricci flatness condition met in the sigma 
models. Note, also, that $R$ in (64) is the scalar curvature of the 2-d 
world sheet of the string. When $R=0$, (64) is the vacuum Einstein equation 
and this agrees with the
results obtained in \cite{poly:book}, \cite{GSW} using the formalism where both 
the 2-d metric and
immersion coordinates are treated dynamically. There this equation is derived 
for flat world sheet. The
crucial difference from their results arises from the scalar curvature of 
the world sheet appearing in
(64). In \cite{poly:book}, it is shown that the divergent Euler characteristic 
term appears in the second order
perturbation and it is absorbed in the renormalization of the Liouville mode.
 We note here that, in the
 formalism used here, this term appears in the first order and is absorbed 
in the background metric
renormalization. Eqn.(64) has a very nice interpretation. First note that as a
 consequence of (64)
$\tilde{R} = (D/2) R$. This makes sense for, when D=2, $\tilde{R}$ evaluated on 
the world sheet must
coincide with the scalar curvature of the world sheet. Using this relation back 
in (64) we can rewrite it as
\begin{eqnarray}
{\tilde{R}}_{\mu\nu} - \frac{1}{2} \tilde{R} h_{\mu\nu} &=& \frac{2-D}{4}R
h_{\mu\nu},
\nonumber 
\end{eqnarray}
which can be interpreted as Einstein equation with a varying cosmological
constant $\Lambda =\frac{2-D}{4}R$.   

We now calculate 1-loop effects for the extrinsic curvature action. The 
coupling constant $\alpha_{0}$ in (20) is asymptotically free in flat 
space-time. So, in this case we will encounter renormalization of both the 
space-time metric and the coupling of the extrinsic curvature action.
As in the case of the N-G action, we start by calculating the second 
variation of the extrinsic geometric action. This, however turns out to 
be rather involved and we sketch the essential steps below.
The first variation of this action given in (33) is, after ignoring the 
total divergence terms and the longitudinal fluctuations terms, (which 
vanish due to the equation of Gauss) can be written as
 
\begin{eqnarray}
\delta S_{R} &=& \int \sqrt{g} ( {\nabla}^{2} H^{i} - 2 H^{2} H^{i} 
+H^{j} H^{j}_{\alpha\beta} H^{i\alpha\beta} \nonumber \\
& & - H^{j} {\tilde{R}}_{\lambda\nu\rho\sigma} {\partial}_{\alpha}X^{\nu} 
{\partial}_{\beta}X^{\rho}g^{\alpha\beta} N^{j\lambda} N^{i\sigma}) 
\xi_{i}(\sigma,\tau) d\sigma d\tau,
\end{eqnarray}
where $\xi_{i}(\sigma,\tau)$ is the normal fluctuation defined in (34).
Starting from the above equation we evaluate its on-shell  
variation. Thus, it is sufficient to evaluate the variation of the terms
within 
the brackets. This is because of the equation of motion (35). It is clear 
that there will be terms that involve operators that couple only normal 
fluctuations and a piece that couples tangential fluctuations to the 
normal fluctuations. It will turn out that these second type of terms 
vanish as a consequence of the equation of motion (35).The calculations 
needed to evaluate the second variation of the extrinsic curvature action 
are lengthy. We give below the key steps.

In addition to the results used in the second variation of the N-G 
action, as well as  
equations (27), (28), (52), and (53), we need the following results:
The variation of $\Gamma^{\gamma}_{\alpha\beta}$:

\begin{eqnarray}
\delta {\Gamma}^{\gamma}_{\alpha\beta} &=& \frac{1}{2} g^{\gamma\delta}( 
\nabla_{\alpha} \nabla_{\beta} + \nabla_{\beta} \nabla_{\alpha}) 
\xi_{\delta} - \frac{1}{2} g^{\gamma\delta} 
R^{\epsilon}\,_{\beta\alpha\delta}\xi_{\epsilon} \nonumber \\
& & - \frac{1}{2} g^{\gamma\delta} R^{\epsilon}\,_{\alpha\beta\delta} 
\xi_{\epsilon} - g^{\gamma\delta} \nabla_{\alpha}(H^{j}_{\delta\beta} 
\xi^{j}) \nonumber \\
& & - g^{\gamma\delta} \nabla_{\beta}(H^{j}_{\alpha\delta} \xi^{j}) + 
g^{\gamma\delta} \nabla_{\delta}(H^{j}_{\alpha\beta} \xi^{j}).
\end{eqnarray}
In (66), $R^{\epsilon}\,_{\beta\alpha\delta}$ is the curvature tensor of 
the world sheet. The longitudinal fluctuations are denoted by $\xi$ with 
greek indices while the normal fluctuations are denoted by $\xi$ with 
latin indices.
Next, the variation of the normal gauge connection $A_{\alpha}^{ij}$:
>From its definition 

\begin{eqnarray}
A_{\alpha}^{ij} = N^{j\nu}(\tilde{\nabla}_{\alpha} N^{i\mu})h_{\mu\nu},
\end{eqnarray}
where $\tilde{\nabla}_{\alpha}$ is defined in (11), it can be verified 
that
 
\begin{eqnarray}
\delta A_{\alpha}^{ij} &=& - F^{ij}_{\alpha\beta} \xi^{\beta} + 
(H^{i}_{\alpha\beta} \nabla^{\beta} \xi^{j} - H^{j}_{\alpha\beta} 
\nabla^{\beta} \xi^{i}) \nonumber \\
& & + \tilde{R}_{\mu\sigma\lambda\rho} N^{i\sigma} 
N^{j\mu}{\partial}_{\alpha}X^{\rho} N^{k\lambda} \xi^{k}. 
\end{eqnarray}
In (86), $F_{\alpha\beta}^{ij}$ is the field strength of 
$A_{\alpha}^{ij}$ defined in (16).

We now give the results for the variation of each of the terms in (65).

\begin{eqnarray}
\delta (\nabla^{2} H^{i}) &=& (\nabla_{\gamma}(\nabla^{2} H^{i}))  
\xi^{\gamma} + \frac{1}{2} \nabla^{2}({\bf O}^{i}_{j} \xi^{j}) + 2 ( 
\nabla^{\alpha} \nabla^{\beta} H^{i})H^{j}_{\alpha\beta} 
\xi^{j} 
\nonumber \\ 
& & - 2(\nabla^{\alpha} H^{j})[(H^{i}_{\alpha\beta} \nabla^{\beta} 
\xi^{j} - H^{j}_{\alpha\beta} \nabla^{\beta} \xi^{i}) \nonumber \\ 
& & + \tilde{R}_{\mu\sigma\lambda\rho} N^{i\sigma} N^{j\mu} 
{\partial}_{\alpha}X^{\rho} N^{k\lambda} \xi^{k}] \nonumber \\
& & - H^{j} \nabla^{\alpha}[\tilde{R}_{\mu\sigma\lambda\rho} N^{i\sigma} 
N^{j\mu} {\partial}_{\alpha}X^{\rho} N^{k\lambda} \xi^{k} \nonumber \\ 
& & + (H^{i}_{\alpha\beta} \nabla^{\beta} \xi^{j} - 
H^{j}_{\alpha\beta} 
\nabla^{\beta} \xi^{i})] \nonumber \\
& & + 2(\nabla^{\gamma} H^{i}) [ \nabla^{\alpha}(H^{j}_{\alpha\gamma} 
\xi^{j}) - \nabla_{\gamma}(H^{j} \xi^{j}) ].
\end{eqnarray}

The variation of $(H^{2}H^{i})$ is given by
\begin{eqnarray}
\delta (H^{2}H^{i}) &=&  (\nabla^{\alpha}(H^{2} H^{i})) \xi_{\alpha} + 
(\frac{1}{2} H^{2} \delta^{i}_{j} + H^{i} H^{j}){\bf O}^{i}_{j} \xi^{k},
\end{eqnarray}
and
\begin{eqnarray}
\delta (H^{j} H^{j}_{\alpha\beta} H^{i\alpha\beta}) &=& 
(\nabla^{\alpha}(H^{j} H^{j}_{\beta\gamma} H^{i\beta\gamma})) 
\xi_{\alpha} + \frac{1}{2} H^{j}_{\alpha\beta} 
H^{i\alpha\beta}{\bf O}^{j}_{k} \xi^{k}  \nonumber \\
& & + H^{j} H^{i\alpha\beta} {\bf O}^{j}_{k,\alpha\beta} \xi^{k} + 
H^{j} H^{j\,\alpha\beta} {\bf O}^{i}_{k,\alpha\beta} \xi^{k},
\end{eqnarray}
where 
\begin{eqnarray}
{\bf O}^{i}_{k,\alpha\beta}&=&
{\nabla}_{\alpha}{\nabla}_{\beta}{\delta}^i_k +
H^i_{\alpha\gamma}H^{\gamma}_{k\beta} - {\tilde{R}}_{\mu\nu\rho\sigma}
N^{i\mu}N^{\sigma}_k{\partial}_{\alpha}X^{\rho}{\partial}_{\beta}X^
{\nu}. 
\end{eqnarray} 
The above operator ${\bf O}^i_{k,\alpha\beta}$ satifies 
\begin{eqnarray}
{\bf O}^{i}_{k,\alpha\beta} g^{\alpha\beta} &=& {\bf O}^{i}_{k}, 
\nonumber  
\end{eqnarray}
which is defined in (54).

The final variation we need is given below.
\begin{eqnarray}
\delta(g^{\alpha\beta}H^{j}\tilde{R}_{\mu\nu\rho\sigma}
{\partial}_{\alpha}X^{\nu}{\partial}_{\beta}X^{\rho}N^{j\mu}N^{i\sigma}) 
&=& \nabla^{\alpha}(H^{j} 
\tilde{R}_{\lambda\nu\rho\sigma}{\partial}_{\alpha}X^{\nu}
{\partial}_{\beta}X^{\rho}g^{\alpha\beta} 
N^{j\lambda} \nonumber \\
& & N^{i\sigma})\xi_{\alpha} \nonumber \\
& & + H^{j}(\nabla_{\mu}\tilde{R}_{\lambda\nu\rho\sigma})
{\partial}_{\alpha}X^{\nu}
{\partial}_{\beta}X^{\rho}g^{\alpha\beta}N^{j\lambda} \nonumber \\
& & N^{i\sigma}N^{k\mu}\xi^{k} \nonumber \\
& & + \frac{1}{2}\tilde{R}_{\lambda\nu\rho\sigma} 
{\partial}_{\alpha}X^{\nu}
{\partial}_{\beta}X^{\rho}g^{\alpha\beta}
N^{j\lambda}N^{i\sigma}{\bf 
O}^{i}_{k}\xi^{k} \nonumber \\
& & + 
H^{j}\tilde{R}_{\lambda\nu\rho\sigma}
N^{j\lambda}N^{i\sigma}
({\partial}_{\alpha}X^{\rho}N^{k\nu} \nonumber \\ 
& & + {\partial}_{\alpha}X^{\nu}N^{k\rho}) 
\nabla^{\alpha}\xi^{k} \nonumber \\
& & + 2H^{j}(\nabla_{\gamma}H^{i})\nabla^{\gamma}\xi^{j} 
+ 2H^{j}(\nabla_{\gamma}H^{j}) \nonumber \\
& & \nabla^{\gamma}\xi^{i} \nonumber \\
& & - H^{j}(\nabla^{\alpha}H^{i}_{\alpha\gamma})\nabla^{\gamma}\xi^{j} - 
H^{j}(\nabla^{\alpha}H^{j}_{\alpha\gamma}) \nonumber \\
& & \nabla^{\gamma}\xi^{i}.
\end{eqnarray}

Making use of the above results we can  write down the second 
variation of the extrinsic curvature action in background space-time. 
Notice that the terms that couple the normal fluctuations to the 
tangential ones vanish because of the equation of motion (35). We thus 
find for $\delta^{2}S$ the following:
\begin{eqnarray}
\delta^{2}S &=& \frac{1}{\alpha_{0}} \int (\xi^{i} {\bf 
\tilde{O}}^{i}_{k}\xi^{k}) \sqrt{g}\,d\sigma d\tau,
\end{eqnarray}
where

\begin{eqnarray}
{\bf \tilde{O}}^{i}_{k} &=& \frac{1}{2} {\bf O^{i}_{j}O^{j}_{k}} - 
(H^{2}\delta^{i}_{j} + 2H^{i}H^{j}){\bf O}^{j}_{k} + 
H^{j}H^{i\alpha\beta}{\bf O}^{j}_{k,\alpha\beta} \nonumber \\
& & + H^{j}H^{j\alpha\beta}{\bf O}^{i}_{k,\alpha\beta} + 
2(\nabla^{\alpha}\nabla^{\beta}H^{i}) H^{k}_{\alpha\beta} \nonumber \\
& &  -2 \nabla^{\alpha}H^{j}(H^{i}_{\alpha\beta}\nabla^{\beta} \delta^{j}_{k} - 
H^{j}_{\alpha\beta} \nabla^{\beta} \delta^{i}_{k}) \nonumber \\
& & - 2(\nabla^{\alpha}H^{j}) 
\tilde{R}_{\mu\sigma\lambda\rho}N^{i\sigma}N^{j\mu}{\partial}_{\alpha}X^{\rho}
N^{k\lambda} \nonumber \\
& & - 
H^{j}(\nabla^{\alpha}[H^{i}_{\alpha\beta}\nabla^{\beta}\delta^{j}_{k} 
- H^{j}_{\alpha\beta} \nabla^{\beta}\delta^{i}_{k} 
 + \tilde{R}_{\mu\sigma\lambda\rho} N^{i\sigma} 
N^{j\mu}{\partial}_{\alpha}X^{\rho}
N^{k\lambda}]) \nonumber \\
& & + 2 \nabla^{\gamma}H^{i}[\nabla^{\alpha}H^{j}_{\alpha\gamma} + 
H^{j}_{\alpha\gamma}\nabla^{\alpha} - \nabla^{\gamma}H^{j} - 
H^{j}\nabla^{\gamma}]\delta^{j}_{k} \nonumber \\
& & -
H^{j}(\nabla_{\mu}\tilde{R}_{\lambda\nu\rho\sigma}){\partial}_{\alpha}X^{\nu
}
{\partial}_{\beta}X^{\rho}g^{\alpha\beta}N^{j\lambda}N^{k\mu} 
-2H^{j}(\nabla_{\gamma}H^{i})\nabla^{\gamma}\delta^{j}_{k} \nonumber \\
& & - 2H^{j}(\nabla_{\gamma}H^{j})\nabla^{\gamma}\delta^{i}_{k} + 
H^{j}(\nabla^{\alpha}H^{i}_{\alpha\gamma})\nabla^{\gamma}\delta^{j}_{k} 
+ H^{j}(\nabla^{\alpha}H^{j}_{\alpha\gamma}) 
\nabla^{\gamma}\delta^{i}_{k}.
\end{eqnarray}
It is straightforward to add the second variation of the N-G action to 
the above. Again, because of the equation of motion for the combined 
action given in (37) the second variation is simply the sum of the two 
variations.

One loop divergent terms can be readily read off from (75). The bare 
propagator for the normal fluctuations is 
\begin{eqnarray}
\langle \xi^{i}\xi^{j}\rangle &=& \delta^{i}_{j} \frac{1}{p^{4}+  
(2\pi\alpha')^{-1}p^{2}}
\end{eqnarray}
As observed in the case of the N-G action, divergent terms involving 
$A_{\alpha}^{ij}$  
do not arise since they can only appear through 
$Tr(F^{\alpha\beta}F_{\alpha\beta})$ which is a dimension four 
operator.There are two types of divergent terms  which are readily 
computed  to give the following result.

\begin{eqnarray}
\Gamma^{1} &=& -DI\int H^{2} \sqrt{g}\, d\sigma d\tau \nonumber \\
& & - I \int (\frac{1}{2}R\, h_{\nu\rho} - \tilde{R}_{\nu\rho}) 
{\partial}_{\alpha}X^{\nu}{\partial}_{\beta}X^{\rho} g^{\alpha\beta} 
\sqrt{g}\,d\sigma d\tau ,
\end{eqnarray}
where $I$ is as given in (57). Divergent extrinsic curvature terms arise 
from  terms 1 to 4 and 8 in (75), whose sum yields the factor 
$D$ in (77). In writing $\Gamma^{1}$ in the form  (77), we have made 
use of the equation of Gauss given in (58).

It is seen that the first term in (77) can be absorbed in the redefinition 
of the extrinsic curvature coupling $\alpha_{0}$ according as 
\begin{eqnarray}
\alpha_{R} &=& \frac{\alpha_{0}}{1 -  \alpha_{0} DI}.
\end{eqnarray}
Notice that the  factor $D$ appears in the denominator in (78), which is the 
dimensionality of the background space-time, rather than $D-2$ that would 
be expected of a non linear sigma model in which the sigma model fields 
are the $D-2$ normals to the surface. We have thus verified that the 
coupling for the extrinsic curvature action in an arbitrary curved 
background is asymptotically free.  We have avoided using the sigma 
model formalism and worked directly in terms of the immersion coordinates 
$X^{\mu}$
as the only dynamical degrees of the theory. The other 1-loop divergent 
term in (77) can be absorbed, as 
in the case of the N-G action, by renormalizing the background metric 
exactly as before. We have not calculated the finite contributions to the 
partition function in this paper.The simplest case would be to integrate 
over minimal surfaces. We can take minimal surfaces with punctures which 
can be interpreted as locations of instanton quarks \cite{vp}. 
\vspace{0.5cm}

\subsection{SUMMARY}

\vspace{0.5cm}
In this paper we have explored both the intrinsic and extrinsic geometric 
properties of strings immersed in a $D$ dimensional curved space-time. We 
have used the formalism  in which  only the string coordinates $X^{\mu}$ 
are the
dynamical fields, while the metric on the world sheet is induced 
from the string's embedding in space-time. This procedure allows one to 
take full advantage of the equations of Gauss, Codazzi, and Ricci for 
immersed surfaces in a general background space-time. It is also 
established that the extrinsic geometry of the strings plays as 
important a role as the Nambu-Goto term in determining the dynamics of 
the strings. We have studied the classical equations of motion and the 
energy-momentum tensor of strings described by both the N-G and extrinsic 
curvature actions. The equations of motion determine the scalar mean 
curvature of the immersed world sheets. We have derived an expression for 
the covariantly conserved energy-momentum tensor for rigid strings. We 
have also studied 1-loop quantum corrections using the background field 
method. The second variation of the action when calculated on-shell 
couples only the normal fluctuations of the world sheet. The tangential 
fluctuations
( zero modes ) decouple from the normal fluctuation modes. We have 
discussed the renormalization of the coupling constant of the extrinsic
curvature action  
as well as the renormalization of the background metric. 1-loop 
finiteness requires a vacuum Einstein type equation.  

\vspace{2.0cm}

{\bf Acknowledgements}:
\vspace{0.5cm}

We thank R.Capovilla and J.Guven for their comments on the original 
version of the manuscript.     
This work is supported in part by an operating grant from the Natural 
Sciences and Engineering Research Council of Canada.

\newpage

\vspace{0.5cm}
\begin{center}
\appendix{APPENDIX}
\end{center}
\vspace{0.5cm}
>From $T^{\mu\nu}$ defined in (47) we demonstrate below that
${\nabla}_{\mu}T^{\mu\nu}\ =\ 0$, when the equations of
motion (35) are satisfied. A direct calculation shows that  
\begin{eqnarray}
{\nabla}_{\mu}T^{\mu\nu} &=& \int d\sigma d\tau
\frac{{\delta}^D(X-X(\sigma,\tau))}{\sqrt{h}}\sqrt{g}\{ 
[-{\nabla}^2H^i+2H^2H^i-H^jH^i_{\alpha\beta}H^{j\alpha\beta}] \nonumber \\
& &  N^{i\nu}  + H^i{\partial}_{\beta}X^{\nu}[2g^{\alpha\beta}{\nabla}_
{\alpha}H^i-{\nabla}_{\alpha}H^{i\alpha\beta}] \nonumber \\
& & + \frac{1}{2}\left(
({\nabla}^2H^i)N^{i\nu}-({\nabla}_{\alpha}H^i)
H^{i\alpha\beta}{\partial}_{\beta}X^{\nu}\right)\} \nonumber \\
& & - \frac{1}{2}
{\nabla}_{\mu}\int d\sigma d\tau \frac{{\delta}^D(X-X(
\sigma,\tau))}{\sqrt{h}} \sqrt{g}\,
g^{\alpha\beta}({\nabla}_{\beta}H^i)N^{i\mu}{
\partial}_{\alpha}X^{\nu} \nonumber \\
& &  -[{\nabla}_{\mu},
{\nabla}_{\rho}]\int \frac{{\delta}^D(X-X(\sigma,\tau))}
{\sqrt{h}}\sqrt{g}\,g^{\alpha\beta}H^i{\partial}_{\beta}X^{\nu}
{\partial}_{\alpha}X^{\rho}N^{i\mu} \nonumber \\
& &  -\frac{1}{2}{\nabla}_{\rho}{\nabla}_{\mu}\int
\frac{{\delta}^D(X-X(\sigma,\tau))}{\sqrt{h}}\sqrt{g}\,g^{\alpha\beta}
H^i{\partial}_{\beta}X^{\mu} \nonumber \\
& &  (N^{i\nu}{\partial}_{\alpha}X^{\rho} - 
N^{i\rho}{\partial}_{\alpha}X^{\nu}). \hspace{5.0cm} (A-1)
\nonumber
\end{eqnarray}
The last term in $(A-1)$ cancels the second and third terms. Finally
considering the 2-dimensional integral in the commutator term as a
tensor $A^{\nu\rho\mu}(X)$ and applying the rule 
\begin{eqnarray}
[{\nabla}_{\mu},{\nabla}_{\rho}] A^{\nu\rho\mu}
&=& R^{\nu}_{\sigma\mu\rho}A^{\sigma\rho\mu} +
R^{\rho}_{\sigma\mu\rho}A^{\nu\sigma\mu} 
+ R^{\mu}_{\sigma\mu\rho} A^{\nu\rho\mu} \hspace{2.0cm} (A-2).  
\nonumber
\end{eqnarray}  
and taking advantage of the Dirac delta function under the integral,
we can write ${\nabla}_{\mu} T^{\mu\nu}$ as
\begin{eqnarray}
{\nabla}_{\mu}T^{\mu\nu}&=& \int d\sigma d\tau
\frac{{\delta}^D(X-X(\sigma,\tau))}{\sqrt{h}} \sqrt{g}\{
[-{\nabla}^2H^i + 2H^2 H^i \nonumber \\
& &  -H^jH^i_{\alpha\beta}
H^{j\alpha\beta}]N^{i\nu} 
+ H^i[2{\nabla}^{\beta}H^i - {\nabla}_{\alpha}
H^{i\alpha\beta}] {\partial}_{\beta}X^{\nu} \nonumber \\
& & + H^i{\tilde{R}}^{\nu}\,_{\sigma\rho\mu}{\partial}_{\alpha}X^{\sigma}
{\partial}_{\beta}X^{\rho} g^{\alpha\beta} N^{i\mu} \}.\hspace{4.0cm} (A-3)
\nonumber  
\end{eqnarray}
Requiring ${\nabla}_{\mu} T^{\mu\nu} = 0$, we find that
\begin{eqnarray}
({\nabla}^2 H^i - 2H^2H^i + H^j
H^i_{\alpha\beta}H^{j\alpha\beta})N^{i\nu}   \nonumber \\
  - H^i{\tilde{R}}^{\nu}\, 
_{\sigma\rho\mu}{\partial}_{\alpha}X^{\sigma}{\partial}_{\beta}
X^{\rho}g^{\alpha\beta}N^{i\mu}  
- H^i [ 2{\nabla}^{\beta}H^i - {\nabla}_{\alpha}
H^{i\alpha\beta}] {\partial}_{\beta}X^{\nu} &=&  0.  (A-4)
\nonumber 
\end{eqnarray}
Multiplying $(A-4)$ by $N^{k\mu} h_{\mu\nu}$  we find this condition is
the equation of motion (35) while the tangential projection yields the
Codazzi equation (36).

\end{document}